\documentclass[%
 aip,
 amsmath,amssymb,
 reprint,%
]{revtex4-1}

\usepackage{graphicx}
\usepackage{dcolumn}
\usepackage{bm}

\usepackage[utf8]{inputenc}
\usepackage[T1]{fontenc}
\usepackage{mathptmx}

\begin{document}

\preprint{AIP/123-QED}

\title{Observation of quincunx-shaped and dipole-like flatband states in photonic rhombic lattices without band-touching}

\author{Shiqiang Xia}
\affiliation{
Engineering Laboratory for Optoelectronic Technology and Advanced Manufacturing, Henan Normal University, Xinxiang, 453007, China
}%
\affiliation{
The MOE Key Laboratory of Weak-Light Nonlinear Photonics, TEDA Applied Physics Institute and School of Physics, Nankai University, Tianjin 300457, China
}%
\author{Carlo Danieli}
\affiliation{
Center for Theoretical Physics of Complex Systems, Institute for Basic Science (IBS), Daejeon 34126, Republic of Korea
}%
\author{Wenchao Yan}
\author{Denghui Li}
\author{Shiqi Xia}%
\author{Jina Ma}%
\affiliation{
The MOE Key Laboratory of Weak-Light Nonlinear Photonics, TEDA Applied Physics Institute and School of Physics, Nankai University, Tianjin 300457, China
}%
\author{Hai Lu}%
\affiliation{
Engineering Laboratory for Optoelectronic Technology and Advanced Manufacturing, Henan Normal University, Xinxiang, 453007, China
}%
 \author{Daohong Song}%
 \email{songdaohong@nankai.edu.cn}
 \affiliation{
The MOE Key Laboratory of Weak-Light Nonlinear Photonics, TEDA Applied Physics Institute and School of Physics, Nankai University, Tianjin 300457, China
}%
\affiliation{%
Collaborative Innovation Center of Extreme Optics, Shanxi University, Taiyuan, Shanxi 030006, China
}%
 \author{Liqin Tang}%
 \email{tanya@nankai.edu.cn}
 \affiliation{
The MOE Key Laboratory of Weak-Light Nonlinear Photonics, TEDA Applied Physics Institute and School of Physics, Nankai University, Tianjin 300457, China
}%
\author{Sergej Flach}
\affiliation{
Center for Theoretical Physics of Complex Systems, Institute for Basic Science (IBS), Daejeon 34126, Republic of Korea
}%
\author{Zhigang Chen}
\email{zgchen@nankai.edu.cn}
\affiliation{
The MOE Key Laboratory of Weak-Light Nonlinear Photonics, TEDA Applied Physics Institute and School of Physics, Nankai University, Tianjin 300457, China
}%
\affiliation{%
Collaborative Innovation Center of Extreme Optics, Shanxi University, Taiyuan, Shanxi 030006, China
}%
\affiliation{%
Department of Physics and Astronomy, San Francisco State University, San Francisco, California 94132, USA
}%

\date{\today}

\begin{abstract}
We demonstrate experimentally the existence of compact localized states (CLSs) in a quasi-one-dimensional photonic rhombic lattice in presence of two distinct refractive-index gradients (i.e., a driven lattice ribbon) acting as external direct current (dc) electric fields. Such a lattice is composed of an array of periodically arranged evanescently coupled waveguides, which hosts a perfect flatband that touches both remaining dispersive bands when it is not driven. The external driving is realized by modulating the relative writing beam intensity of adjacent waveguides. We find that a y-gradient set perpendicularly to the ribbon preserves the flatband while removing the band-touching. The undriven CLS - which occupies two lattices sites over one unit cell - turns into a quincunx-shaped CLS spanned over two unit cells. Instead, an x-gradient acting parallel to the ribbon yields a Stark ladder of CLS whose spatial profile is unchanged with respect to the undriven case. We notably find that their superposition leads to Bloch-like oscillations in momentum space.
\end{abstract}
\maketitle
\section{Introduction}

Flatband geometries \cite{PhysRevLett.62.1201,Mielke_1991,PhysRevLett.69.1608,PhysRevB.78.125104,PhysRevB.81.014421,PhysRevA.82.041402,PhysRevA.87.023614,huang2016localization,PhysRevB.96.161104,PhysRevB.96.155137,leykam2018perspective,leykam2018artificial} have attracted great interest in recent years due to the existence of at least one completely dispersionless band in their energy spectrum which bring new perspectives to the study of various fascinating phenomena, including fractional quantum Hall effect \cite{PhysRevLett.106.236802,PhysRevLett.106.236803,PhysRevLett.106.236804,PhysRevLett.107.146803}, inverse Anderson localization \cite{
PhysRevLett.96.126401,PhysRevB.82.104209,PhysRevB.88.224203,PhysRevLett.113.236403,PhysRevB.91.235134,PhysRevLett.116.066402}, conservative PT-symmetric compact solutions \cite{Yulin:13,PhysRevA.92.052103,PhysRevA.92.063813,PhysRevLett.120.093901,PhysRevB.96.064305,Ge:18}, and nonlinear compact breathers \cite{PhysRevE.92.032912,PhysRevB.94.144302,PhysRevA.96.063838,DanieliLowTemperaturePhysics}. Destructive interference is the essence of a flatband existence, and the associated eigenmodes are compact in real space - hence dubbed compact localized states (CLSs). The robustness of the spatial compactness of such CLSs has been observed in various models, particularly in artificial Lieb \cite{PhysRevLett.114.245503,PhysRevLett.114.245504,Xia:16,PhysRevLett.116.183902,Taiee1500854}, Kagome \cite{Vicencio_2013,Zong:16}, and rhombic lattices \cite{Mukherjee:15,Mukherjee:17}. More recently, the interplay between flatband and external driving field has been investigated, leading to intriguing phenomena such as topological flatband insulators \cite{PhysRevB.82.085310,BergholtzInternationalJournalofModernPhysics,tang2014strain,PhysRevLett.120.097401}, unconventional Landau–Zener Bloch oscillations \cite{PhysRevLett.116.245301,PhysRevLett.118.175301,long2017topological}, and magnetic field-induced Aharonov–Bohm caging \cite{Longhi:14,PhysRevLett.121.075502,kremer2018non,PhysRevA.99.013826}. However, a fundamental question remains elusive: how do the CLSs change in the presence of external fields? Mukherjee and Thomson showed that the CLSs in a quasi-one-dimensional rhombic lattice are robust in the presence of external driving potential \cite{Mukherjee:17}, while Khomeriki and Flach predicted that CLSs are also robust in the presence of direct current (dc) electric and magnetic fields \cite{PhysRevLett.116.245301}. However, in a two-dimensional dice lattice, Kolovsky et.al. showed that CLSs stop being compact in presence of a dc electric field, and instead they turn into exponentially (super-exponentially) localized in the perpendicular (parallel) direction of the field \cite{PhysRevB.97.045120}. Despite those theoretical studies, the intrinsic mechanism played by external fields in flatband systems is still unclear, and in many cases not supported by experimental observation.

\begin{figure*}
\centering
\includegraphics[width=0.9\textwidth]{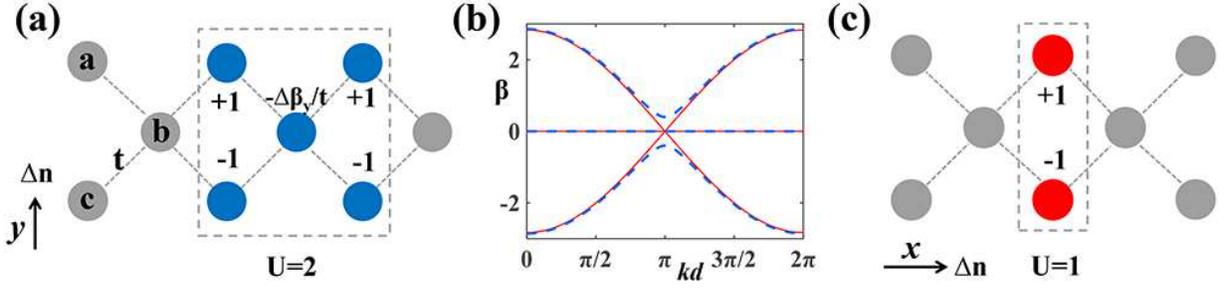}
\caption{\label{fig:1} (a) Schematic diagram of a rhombic lattice with y-gradient. Each unit cell consists of three sites (a, b, and c) and a quincunx-shaped U = 2 CLS structure is marked by a dashed square. (b) Band structure of the rhombic lattices for coupling coefficient $t=1$, with a y-gradient $\Delta \beta _{y}=0.4$ (dashed blue lines). Solid red lines represent spectrum of uniform lattices without gradient ($\Delta \beta _{y}=0$). (c) U = 1 CLS structure (shown in dashed rectangle) of rhombic lattices with x-gradient. Sites with nonzero amplitude in (a) and (c) are denoted by blue and red circles, the numbers and sign near the sites represent the amplitudes and phases, respectively.}
\end{figure*}
In this work, we experimentally demonstrate for the first time the existence of quincunx-shaped CLSs in a quasi-one-dimensional photonic rhombic lattice without band-touching formed by an array of evanescently coupled waveguides in presence of external dc fields. Using a continuous-wave (cw) laser writing technique, we  experimentally establish finite-sized photonic rhombic lattices and introduce refractive index gradients perpendicular (y-gradient) and parallel (x-gradient) to the ribbon. A periodic distribution of the refractive index plays a role of the periodic potential, and the refractive index gradient is the optical counterpart of an external driven force in a quantum system. On the one hand, we show that the y-gradient does not lift the flatband at zero energy, but it removes the touching with the dispersive bands, introducing a gapped band structure. The associated CLSs exhibit a quincunx pattern which spans over two unit cells of the lattices – a shape which is preserved during the propagation along the waveguides. Such CLSs represent a new type of flatband modes in accelerated flatband lattices whose intensity and phase structure are quite different from conventional ones. In addition, the existence of such CLSs arises from the interplay between the flatband and the external dc electric field, which may provide insight to dynamics of flatband states under other external fields such as a magnetic field. On the other hand, in the presence of an x-gradient, the photonic eigenstates form an optically equivalent Wannier-Stark ladder by triplets of eigenvalues - one of which is the CLSs energy - equispaced along the real axis by a shifting factor proportional to the field $\Delta\beta_{x}$. The associated CLSs exhibit unaltered spatial profile with respect to an undriven rhombic chain. However, we find that superimposed CLSs leads to unconventional Bloch-like oscillations in momentum space during the propagation.

\section{Materials and methods}

\begin{figure*}[htb]
\centering
\includegraphics[width=0.9\textwidth]{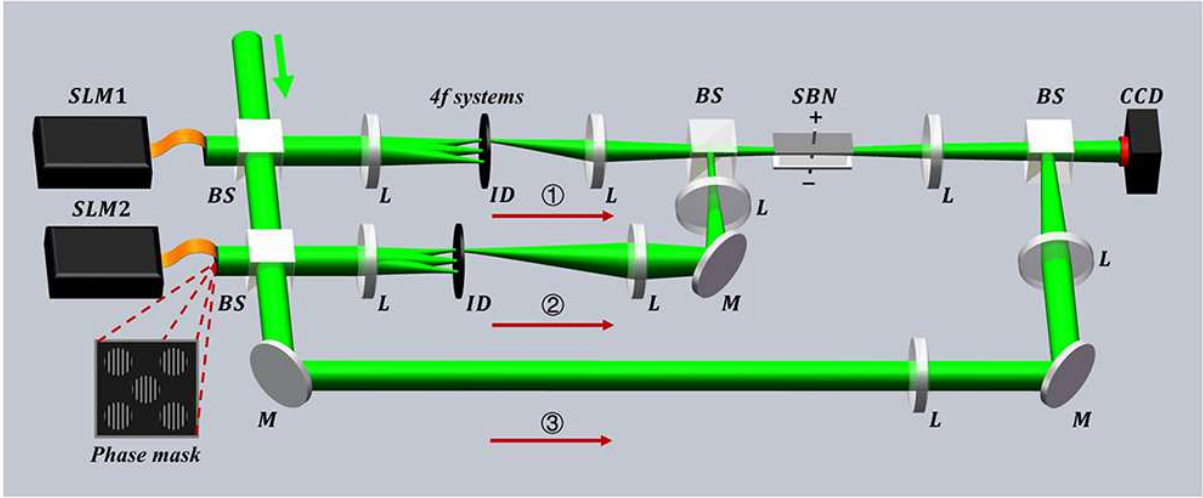}
\caption{\label{fig:2} Experimental setup for observation of flatband CLSs in optically induced driven photonic rhombic lattices. SLM: spatial light modulator; ID: iris diaphragm; SBN, strontium barium niobite; L: lens; M: mirror; BS: beam splitter. Phase mask shown here represents the phase pattern used to generate a quincunx-shaped probe beam. Red arrow 1 shows site-to-site writing of photonic lattices with a cw laser in a nonlinear crystal, and arrows 2, 3 show, respectively, the probe beam, and the interfering beam for measuring the output phase structure.}
\end{figure*}

A photonic rhombic lattice consists of three sites (a, b, and c) per unit cell [Fig.~\ref{fig:1}(a)]. Such a geometry has been previously used to theoretically and experimentally study various interesting effects mentioned above \cite{PhysRevB.88.224203,PhysRevLett.113.236403,PhysRevB.91.235134,PhysRevB.94.144302,Mukherjee:15,Mukherjee:17,PhysRevLett.116.245301,Longhi:14,PhysRevLett.121.075502,kremer2018non,PhysRevA.99.013826}. Although being a simple flatband geometry, it allows to clarify the effects of the refractive index gradient fields on the model’s CLSs. We investigate this system in the tight binding approximation, whose Hamiltonian can be written as \cite{Mukherjee:15,Mukherjee:17,PhysRevLett.116.245301,Longhi:14,PhysRevLett.121.075502,kremer2018non,PhysRevA.99.013826}:

\begin{eqnarray}
{H_t} &&= \sum\limits_n {t\left( {b_n^\dag {a_n} + b_n^\dag {a_{n - 1}} + b_n^\dag {c_n} + b_n^\dag {c_{n - 1}} + H.c.} \right)} \nonumber \\
{H_y} &&= \sum\limits_n {\Delta {\beta _y}\left( {a_n^\dag {a_n} - c_n^\dag {c_n}} \right)}\nonumber \\
{H_x} &&= \sum\limits_n {2n\Delta {\beta _x}\left( {a_n^\dag {a_n} + b_n^\dag {b_n} + c_n^\dag {c_n}} \right)}  - \Delta {\beta _x}b_n^\dag {b_n}
\label{eq:one}
\end{eqnarray}
where $a_{n}^{\dagger}$, $b_{n}^{\dagger}$, $c_{n}^{\dagger}$ and $a_{n}$, $b_{n}$, $c_{n}$ are the creation and annihilation operators in the n-th unit cell on the a, b and c sites, respectively. Here $\Delta\beta_{x}$ and $\Delta\beta_{y}$ denote the wave-number spacing between adjacent waveguides and define the applied linear refractive index gradient $\Delta{n}$ parallel and perpendicular to the lattices. The effective propagation constant (or the on-site energy) of the waveguides is determined by the index gradient strength and its direction. In the presence of y-gradient field, we assume that sites a, b and c of the same unit cell have onsite energy difference $\Delta\beta_{y}$ along y direction. And, in the presence of x-gradient, sites a and c of the n-th unit cell have the same effective propagation constant which is shifted by $2\Delta\beta_{x}$ compared to the same sites of the (n+1)th and the (n-1)th unit cell, respectively.

In the absence of external fields ($\Delta \beta _{x}=0$, $\Delta \beta _{y}=0$), the Bloch representation yields the dispersion relation of the uniform lattices, which is made of three spectral bands: a completely degenerated flatband at zero energy $\beta _{flat}=0$ located between two dispersive bands $\beta _{\pm }= \pm \sqrt{8t^{2}\sin^{2}\left ( kd/ 2\right )}$, here $d$ is the lattice constant. These energy bands can be observed in Fig.~\ref{fig:1}(b) for $t=1$ (solid red lines). All three bands touch at a high symmetric point of the first Brillouin zone. The irreducible CLS of the flatband is sketched in Fig.~\ref{fig:1}(c) which only occupies a and c sites within one unit cell, with equal amplitude and opposite phase, ensuring destructive interference in the neighboring b sites. Following the typical characterization of CLSs in one-dimensional settings by the integer number U of unit cells they occupied \cite{Flach_2014}, the CLS shown in Fig. 1(c) is of class U = 1. When a y-gradient is applied ($\Delta \beta _{x}=0$, $\Delta \beta _{y}\neq0$), the rhombic chain Eq.~(\ref{eq:one}) is still translation invariant with the following dispersion relations:$\beta _{flat}=0$, $\beta _{\pm }= \pm \sqrt{\left (\Delta \beta _{y} \right )^{2}+8t^{2}\sin^{2}\left ( kd/ 2\right )}$. As shown in Fig. 1(b) for $t=1$ and  $\Delta \beta _{y}=0.4$  (dashed blue lines), band-touching vanishes and two symmetric gaps open between the central flatband and two dispersive bands. Since the y-gradient breaks the local symmetry perpendicular to the rhombic lattices, the conventional U = 1 CLSs no longer exist. Instead, the CLS for $\beta _{flat}=0$ located at the n-th unit cell has the spatial profile
\begin{equation}
\Psi _{\beta = 0}= \begin{pmatrix}
\left ( a_{n}+a_{n-1}\right ) \\
 -\left ( \Delta \beta_{y} /t\right )b_{n}\\
 -\left ( c_{n}+c_{n-1}\right )
\end{pmatrix}
\label{eq:two}
\end{equation}
This compact state consists of five non-zero amplitude sites (two a, one b and two c sites) arranged in an x-shaped profile coined quincunx - as shown within the square of Fig.~\ref{fig:1}(a). We can see that all the a and c sites in this new-type CLS (covering two unit cells) have the same amplitude but $\pi$-phase difference, ensuring destructive interference on the neighboring b sites. The excitation $-\Delta \beta_{y} /t$ on the central b site of the CLS linearly depends on the strength $\Delta \beta_{y}$ and direction of index gradient, while being inversely proportional to the hopping strength $t$. The amplitude and phase of this central site are crucial to balance the propagation constants on the a and c sites, which are otherwise misbalanced by the y-gradient. Consequently, the CLS in Eq.~(\ref{eq:two}) does not follow from a superposition of the U = 1 flatband states of uniform rhombic lattices, and it is of class U = 2 since it occupies two unit cells.

In the presence of an x-gradient ($\Delta \beta _{x}\neq0$, $\Delta \beta _{y}=0$), the rhombic chain is not translationally invariant, and as a consequence the Bloch representation is no longer applicable. In this case, the energy spectrum consists of a triplet ladder. This ladder is obtained from one triplet which is shifted indefinitely and equidistantly along the real axis with a shift proportional to the strength of the dc field $\Delta\beta_{x}$. Importantly, this parallel dc field does not lift nor deform the class U = 1 CLSs possessed by the uniform rhombic lattices shown in Fig.~\ref{fig:1}(c), but it turns the CLSs energies to be unit-cell dependent $\beta _{flat}=2 \nu \Delta\beta_{x}$, with $\nu$ being an integer. Indeed, since the refractive index gradient field is parallel to the ribbon, the Hamiltonian Eq.~(\ref{eq:one}) remains invariant under the symmetry $a\left ( c\right )\rightarrow c\left ( a\right )$, ensuring the destructive interference of two waves with same amplitude and $\pi$-phase difference located in the a and c sites of one unit cell. Consequently, the energy spectrum is a Stark ladder of triplets of eigenenergies, where the CLSs energy $ 2 \nu \Delta \beta_x$ is part of a triplet completed by the two energies $2\nu \Delta \beta_x$ and $(2 \nu - 1) \Delta \beta_x$. The parallel field alone leads to Bloch oscillations of the dispersive states with oscillation length proportional to $L \sim 2 D_d / \Delta \beta_x$, with $D_d$ the width of the undriven bands. When the parallel field is applied simultaneously with the perpendicular field instead two scenarios emerge: (i) for $\Delta \beta_x \ll G$ (with $G$ the band gap between the dispersive bands opened by the vertical field) the states of different bands do not mix and show oscillation length $L/2$ for the dispersive bands, and an oscillation length of one unit cell for the flat band; (ii) for $\Delta \beta_x \geq G$ the states of different bands mix, leading to a novel oscillation length $L_2 = (2D_d+2G)/ \Delta \beta_x$.

\begin{figure*}[htb]
\centering
\includegraphics[width=0.9\textwidth]{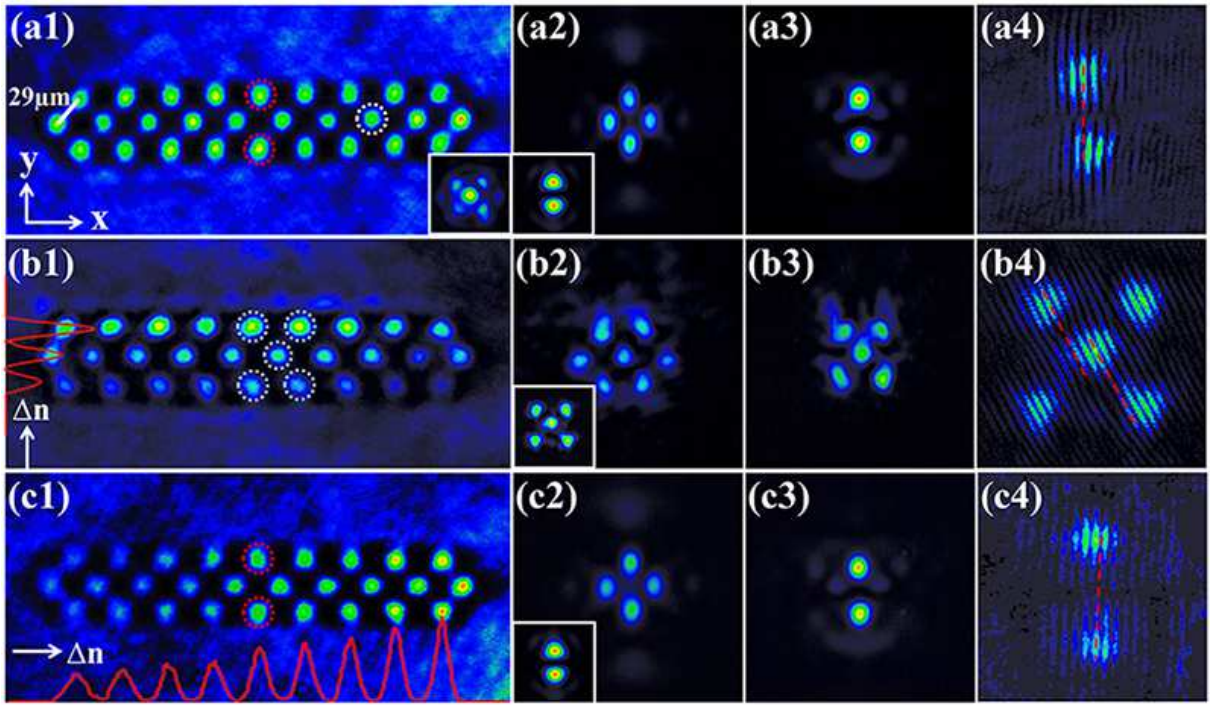}
\caption{\label{fig:3} Experimental observation of CLSs in cw-laser writing photonic rhombic lattices. First column: (a1) A uniform lattice without gradient, (b1, c1) a driven lattice with y-gradient and x-gradient respectively, where the dashed circles indicate the initially excited sites (at z = 0) and inset in (a1) shows discrete diffraction by exciting a single b site (marked by the white dashed circle) with a Gaussian beam. Red line in (b1) indicates the average peak intensity profile of the three sublattices and that in (c1) represents the intensity profile of sublattice in the first row, which indicates different gradient fields of the lattices; Second column: in-phase output at the back facet of the crystal: discrete diffraction appears. The insets show the intensity patterns of the input probe beams. Third column: out-of-phase output at the back facet of the crystal: the CLSs stay in the same position as the input without diffraction. Fourth column: corresponding interferograms. The middle row corresponds to the quincunx-shaped U = 2 flatband states.}
\end{figure*}

To demonstrate the flatband CLSs experimentally, we use a cw-laser writing technique to establish the finite-sized photonic rhombic lattices with desired refractive index gradient (Fig.~\ref{fig:2}). The technique relies on site-to-site inducing or writing waveguides in a nonlinear photorefractive crystal (SBN), and it has already been successfully used in our previous work to design two-dimensional Lieb lattices \cite{PhysRevLett.121.263902}. A laser beam ($\lambda$ = 532 nm) is used to illuminate a phase-only spatial light modulator (SLM1), which creates a quasi-nondiffracting writing beam propagation through a 10-mm-long crystal with reconfigurable input positions (beam path 1). Owing to the noninstantaneous self-focusing nonlinearity, all waveguides remain intact within the one-by-one writing and data acquisition period. Moreover, refractive index gradient fields can be introduced and tuned by varying the writing beam intensity or the bias field of the modulation. To generate the input probe beams, another SLM (SLM2) is used so that we can control the intensity pattern, as well as the phase structure of the probe beam (beam path 2). We simultaneously encode the amplitude and phase information onto the SLM by designing a hologram (phase mask) consisting of several phase gratings arranged in a dipole-like or quincunx-shaped structure. An extraordinarily polarized quasi-plane wave is sent to the SLM, and the first order of the diffracted light whose intensity distribution has a desired pattern is imaged to the facet of the crystal as a probe beam. The size of each spot of the probe beam is controlled by the imaging lens, and the relative intensity of each spot can be tuned by adjusting the input beam width. At the same time, the phase structure is controlled by changing the relative locations of the gratings. Once the probe beam is shaped for the desired exciting condition, it is sent into the induced rhombic lattices, and the output pattern is monitored at the back facet of the crystal. Beam path 3 is the interfering beam for measuring the output phase structure.

\section{Experimental results and analysis}

Experimental results are shown in Fig.~\ref{fig:3}. To visualize the induced rhombic lattices, we illuminate a weak extraordinarily polarized quasi-plane wave to probe the waveguides induced in the crystal. At the back facet of the crystal, one can clearly find that the otherwise uniform probe beam is guided into each lattices site (see first column of Fig.~\ref{fig:3}). In our work, we first establish uniform rhombic waveguide structure without refractive index gradient. Figure~\ref{fig:3}(a1) shows the finite photonic lattices with nine unit cells. The lattices spacing is about 29 $\mu$m. To probe the energy coupling property of the lattices and diffractionless feature of CLSs, we show the propagation dynamics of a Gaussian beam and the fundamental localized flat-band mode. As can be seen in the inset of Fig.~\ref{fig:3}(a1), after propagating through the 10-mm crystal, a single-site excitation leads to discrete diffraction with the energy coupling mainly to the nearest waveguides. The coupling therefore occurs mainly among nearest-neighboring waveguides, which satisfies the tight-binding approximation Eq.~(\ref{eq:one}). Then, a dipole-like beam [inset of Fig.~\ref{fig:3}(a2)] that excites sites a and c of one unit cell is set as input. If the dipole-like beam is out-of-phase, the probe beam stays well localized in the initially excited lattices sites, experiencing no diffraction [Fig.~\ref{fig:3}(a3)] due to the excitation of the U = 1 CLS shown in Fig.~\ref{fig:1}(c). Moreover, the phase measurement obtained by interfering the output with an inclined plane wave further confirms that the initial out-of-phase structure is well preserved [Fig.~\ref{fig:3}(a4)]. However, if the input dipole-like beam is initially in-phase, the CLS is not excited and the output displays discrete diffraction with beam intensity evolving into nearby lattices sites [Fig.~\ref{fig:3}(a2)].

\begin{figure*}[htb]
\centering
\includegraphics[width=0.9\textwidth]{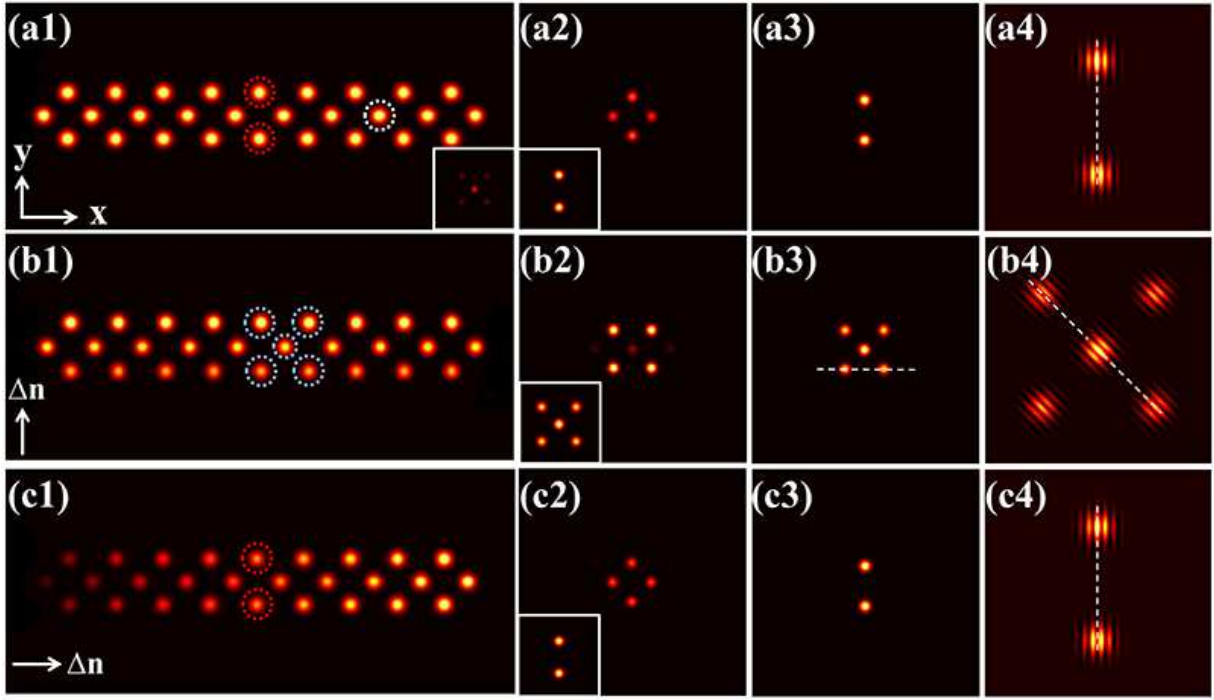}
\caption{\label{fig:4} Simulation results of CLSs in photonic rhombic lattices corresponding to Fig. 3. Each output image is normalized so that the total intensity is 1.}
\end{figure*}
Next, a y-gradient field ($\Delta \beta _{y}\neq0$) is introduced in the uniform rhombic lattices by fine tuning the relative writing beam intensities corresponding to the three sublattices. The measured peak intensity profile indicated with the solid red line is shown in Fig.~\ref{fig:3}(b1). The intensities of the sublattices increase by about $30\%$ perpendicular to the ribbon. Experimentally, such a distinct intensity difference ensures that we can easily observe the effect of y-gradient field. As mentioned above, U = 2 CLSs occupy two unit cells and have unique phase and intensity structure. To match the flatband mode, we launch a probe beam [inset of Fig.~\ref{fig:3}(b2)] which consists of five spots and set the upper two spots in phase to excite the a sites, while the lower three spots to excite the b and c sites with $\pi$ phase difference with a sites. The amplitude of the central b site should match the flatband mode shown in Fig.~\ref{fig:1}(a). For simplicity, we set the intensity of the central spot 1.2 higher than that of the rest four spots. Figure.~\ref{fig:3}(b3) shows the output intensity pattern, and no significant tunneling of light into the surrounding sites emerges. Interference pattern [Fig.~\ref{fig:3}(b4)] also reveals that the phase structure is well preserved, indicating that the input beam stays localized during propagation. On the contrary, when we set the central spot of the input beam with $\pi$ phase shift, i.e., maintain the a and the c sites with out-of-phase structure and keep other experimental parameters fixed, a completely different behavior is observed. The input beam cannot localize and the light energy couples to nearby waveguides [Fig.~\ref{fig:3}(b2)]. Therefore, our study reveals that such a U = 2 CLS is highly sensitive to the phase alterations at the extremities of the quincunx.

Finally, we study the transport property of CLSs in waveguide arrays with x-gradient field ($\Delta \beta _{x}\neq0$). The red line in Fig.~\ref{fig:3}(c1) shows the measured intensity profile of sublattice a, which increases by about $15\%$ in each unit cell along x axis. As discussed above, the U = 1 CLS is still a compact solution of Eq.~(\ref{eq:one}). We use the same probe beam and experimental parameters as that used in uniform lattices, and the results are shown in the third row of Fig. 3.  If the two spots of the dipole-like beam have $\pi$ phase difference, the probe beam is trapped at the initially excited sites as shown in Fig.~\ref{fig:3}(c3). Instead if the two spots of the dipole-like beam have equal phase, discrete diffraction pattern is obtained [Fig.~\ref{fig:3}(c2)]. It should be noted that we get similar discrete diffraction patterns in Figs. ~\ref{fig:3}(a2) and (c2) mainly due to the weak coupling of the lattices and limited propagation length. In fact, the observation of U = 1 CLSs in rhombic lattices with x-gradient was first attempted in curved arrays fabricated by ultrafast laser inscription technique \cite{Mukherjee:17}. In that case, the external horizontal drivings were realized by modulating the paths of the waveguides. We want to mention that though we observe the diffraction suppression of both class U = 1 and U = 2 CLSs in both parallel (x-gradient) and perpendicular (y-gradient) driving of the lattices ribbon, our numerical analysis shows that the dynamics of these two types of CLSs are quite different.

\begin{figure*}[htb]
\centering
\includegraphics[width=0.9\textwidth]{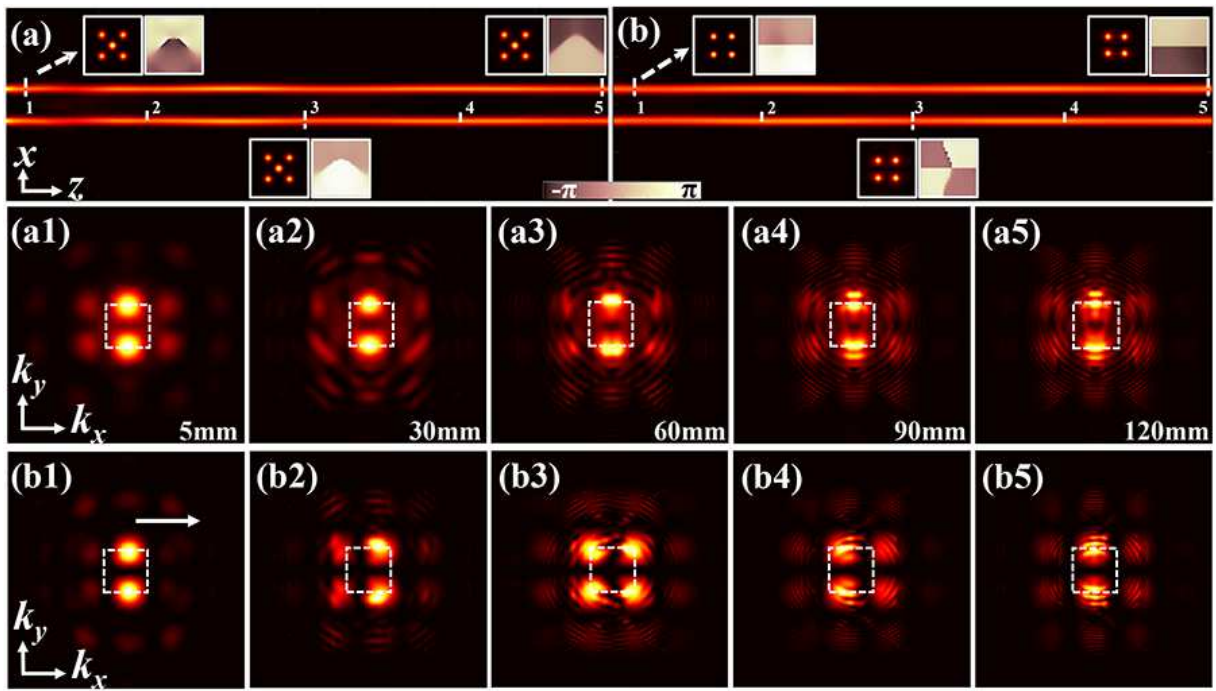}
\caption{\label{fig:5} First row: Side view of beam propagation ($z = 120$ mm) corresponding to (a) one U = 2 CLS along the direction of the dashed white line in Fig. 4(b3), and (b) a superposition of two neighboring U = 1 CLSs. The insets show the output intensity pattern and phase structure for $z= 5, 60,120$ mm, revealing the robust localization of CLSs in real space. Second row: Fourier spectra of U = 2 CLSs for different propagation length ($z= 5, 30, 60, 90, 120$ mm) under a y-gradient field. Third row: Fourier spectra of two superimposed U = 1 CLSs, revealing oscillation under an x-gradient field. The white arrow in (b1) indicates the acceleration direction of the spectra, while the white dashed squares indicate the boundary of Brillouin zone.}
\end{figure*}

We numerically calculate the variation of light intensity along the propagation direction using coupled-mode equations Eq.~(\ref{eq:one}) setting the coupling constant t is 40 $m^{-1}$ while the wave-number spacings $\Delta\beta_{x}$ and $\Delta\beta_{y}$ are set to 25 $m^{-1}$ and 50 $m^{-1}$ respectively, which are similar to the experimental parameters. Results are shown in Fig.~\ref{fig:4}, where it can be seen that our experimental results are in good agreement with the simulation results. In particular, it can be noticed in Fig.~\ref{fig:4}(b2) that the energy of the central spot of the quincunx-shaped excitation shown in the inset decays very fast with the energy coupling to nearby sites, in a striking contrast to the robustness of CLS in Fig.~\ref{fig:4}(b3). Moreover, note that compare to discrete diffraction shown in Fig.~\ref{fig:4}(a2), the output in Fig.~\ref{fig:4}(c2) will experience Bloch oscillation if the propagation length is long enough as predicted in Ref. [41], which exactly reflecting the effect of x-gradient. Intuitively, the presence of both y-gradient and x-gradient have no influence on dynamics of flatband CLSs as they always exhibit robust localization in real space as illustrated in Figs.~\ref{fig:3}(b3) and ~\ref{fig:3}(c3). Nevertheless, we will show in Fig.~\ref{fig:5} that the U = 1 and the U = 2 CLSs show different dynamics in momentum space after longer propagation distance.

Since it is still a challenge to experimentally study the propagation dynamics of the CLSs for long distance due to limitations on the crystal length, we numerically simulate the behavior of both U = 1 and U = 2 CLSs with the propagation distance up to $z=120$ mm. In order to trace the evolution of Fourier spectra and phase structures, we use the paraxial wave equation \cite{PhysRevLett.121.263902} which describes the wave dynamics close to the actual experiment. Moreover, for sake of comparison between U = 1 and U = 2 CLSs, we use a probe beam composed of four spots which can be seen as a superposition of two probe beams in Fig.~\ref{fig:4}(c2) when studying the dynamics of U = 1 CLSs, recalling that the two U = 1 CLSs have energy difference of $2\Delta\beta_{x}$. Figures. ~\ref{fig:5}(a) and ~\ref{fig:5}(b) show the side view of beam propagation corresponding to U = 2 and U = 1 CLSs, respectively. These plots clearly reveal the undistorted transmission of U = 2 CLSs under a y-gradient and U = 1 CLSs under an x-gradient. The insets of Figs.~\ref{fig:5}(a) and ~\ref{fig:5}(b) show the output intensity patterns for different propagation distances corresponding to $z=0$, 60 and 120 mm, confirming the robust localization in real space. Nevertheless, when tracing the evolution of the Fourier spectrum and phase structure, we find that the two classes of CLSs exhibit different behaviors. The middle row of Fig.~\ref{fig:5} shows the Fourier spectra for $z= 5, 30, 60, 90, 120$ mm, corresponding to Fig.~\ref{fig:5}(a). After propagating to $z=5$ mm [Fig.~\ref{fig:5}(a1)], the probe beam ceases to be a stable CLS of the y-gradient rhombic lattices. It can be clearly seen that the spectra are substantially invariant during the long propagation distance. Moreover, the phase structures in real space are also well preserved as shown in Fig.~\ref{fig:5}(a). However, for U = 1 CLSs under an x-gradient, the Fourier spectra shift along the direction of the driving field in Fig.~\ref{fig:5}(b1) as the propagation distance z increases. The bottom row shows the Fourier spectra corresponding to Fig.~\ref{fig:5}(b). At $z=30$ mm [Fig.~\ref{fig:5}(b2)], the beam just experiences its first Bragg reflection and after propagating to $z=60$ mm [Fig.~\ref{fig:5}(b3)], the spectrum transforms to the boundary of Brillouin zone. Additionally, the phase structure of the output becomes a checkerboard pattern at $z=60$ mm which is the characteristic feature of the Bloch mode at the boundary of the Brillouin zone. The CLS completes a full Bloch-like oscillation after propagating to $z=120$ mm, i.e., both spectrum and phase structure recover the initial state as shown in Fig.~\ref{fig:5}(b5). Then, the beam will accelerate again and experience oscillation periodically.

\section{Conclusion}
In conclusion, we have proposed and demonstrated different types of CLSs in photonic rhombic lattices driven by external dc fields along two different directions. Asymmetric features of the CLSs with respect to the driving potential applied have been observed. On the one hand, when a y-gradient is set perpendicularly to the ribbon, the flatband is preserved although lifting the band-touching with the dispersive bands. In this case the undriven class U = 1 CLSs reform dramatically turning into class U = 2 quincunx-shaped CLSs. Such novel CLSs cannot be obtained from a simple superposition of the class U = 1 flatband states of the undriven rhombic lattices, since the y-gradient breaks the local symmetry and lifts the flatband-touching. On the other hand, the undriven class U = 1 dipole-like CLSs preserve with an x-gradient parallel to the ribbon, although their energies are equidistantly distributed Stark ladders along the real axis. Interestingly, these two distinct CLSs exhibit different oscillation properties in momentum space, namely no oscillations emerge for the former U = 2 CLSs, while the superposition of neighboring U = 1 CLSs are characterized by novel Bloch-like oscillations during propagation. Our work paves the way to experimentally achieve non-trivial Landau-Zener Bloch oscillations in lattices when both parallel and perpendicular fields are simultaneously applied, as well as oscillations emerging from the interplay of external dc fields with a uniform magnetic field \cite{PhysRevLett.116.245301}. Possible extensions of the discussed results beyond the considered rhombic lattices include two-dimensional flatband geometries, where it has been recently shown that dc fields applied along specific directions lead to Wannier-Stark ladders of edge states, as well as super-exponentially localized states \cite{PhysRevB.97.045120}. This may also open the avenue for a plethora of further interesting experiments in two-dimensional flatband lattices hunting for fractional charge transport and topological matter, among others.  Lastly, these results are general and applicable to other flatband systems, such as electrons in crystals and ultracold atoms in optical lattices.

\begin{acknowledgments}
This work is supported by the National Key R$\&$D Program of China (2017YFA0303800), the NSFC (11922408, 11704102, 91750204 and 11674180), 111 Project (No. B07013), Doctoral Starting up Foundation (qd16169), Science Foundation for Young Scientists of Henan Normal University (20180051), and the Institute for Basic Science (IBS) -Project Code No. IBS- R024-D1.
\end{acknowledgments}

\nocite{*}
\bibliography{reff}

\end{document}